\documentclass[12pt]{article}
\usepackage{epsfig}
\usepackage{axodraw}
\usepackage{amsfonts}
\usepackage{graphics}

\begin{document}

\begin{center}
{\Large \bf Transverse positron polarization in the $\mu^+ \to e^+ \bar {\nu_{\mu}} \nu_e$ decay in SM.}
\\ \vspace*{5mm} V.V.Braguta\footnote {Electronic address: braguta@mail.ru} \\
{\small \it Institute for High Energy Physics, Protvino, Russia }
\end{center}

\vspace*{0.5cm}

\begin{abstract}
In this paper transverse positron polarization in the $\mu^+ \to e^+ \bar {\nu_{\mu}} \nu_e$ decay in the
framework of SM is considered. It is shown that  the final state interaction effect
leads to nonzero transverse polarization. Numerical value of the considered effect
proved to be negligible. Thus SM contribution to the transverse positron polarization in
the $\mu^+ \to e^+ \bar {\nu_{\mu}} \nu_e$ decay will not be the obstacle to a new physics
searches.
\end{abstract}

The study of muon decay $\mu^+ \to e^+ \bar {\nu_{\mu}} \nu_e$ can give
valuable knowledge about lepton sector of weak interactions. The Standard Model
successfully describes this decay by vector interaction of four left-handed fermions.
To go beyond Standard Model(SM) one introduces into the lagrangian scalar, vector and tensor interactions of
right- and left-handed particles \cite{Fetscher:1986uj}. All these interactions can be parameterized
by 10 complex constants or  one common unphysical
phase with nineteen independent real parameters. If some of this
parameters are nonzero there exists $CP$--violation in purely leptonic
decays.

One possible way to study the phenomenon of $CP$--violation in the
muon decay $\mu^+ \to e^+ \bar {\nu_{\mu}} \nu_e$ is the measurement of transverse
positron polarization that proved to be sensitive to $CP$--violation\cite{kinoshita}. Recently
the measurement of this observable was improved \cite{Danneberg:2005xv} and new upper bound
on the parameters of $CP$--violation has been obtained. The averaged value of
transverse positron polarization obtained at this experiment is
\begin{eqnarray}
<P_{T2}> = (-3.7 \pm 7.7 \pm 3.4) \times 10^{-3}
\end{eqnarray}
Further improvement of the accuracy can lead to the discovery of
$CP$--violation in leptonic decay or can put more strict bounds on parameters
of different SM extensions.

In order to study possible SM extensions by the measurement of the transverse positron polarization
one needs to find SM contribution to the observable. Transverse muon polarization in $K^+ \to
\pi^0 \mu \nu_{\mu}$\cite{Zhitnitsky:he, Efrosinin:2000yv}, $K^+ \to \mu \nu_{\mu} \gamma$\cite{Braguta:2002gz, Rogalev:2001wx},
$ K^0 \to \pi^- \mu^+ \nu$\cite{Okun:1967ww}
decays, $T$-odd correlation in
$K^+ \to \pi^0 \mu \nu_{\mu} \gamma$\cite{Braguta:2001nz} decay etc. are the examples of similar physical
observables that are very sensitive to the effect of $CP$--violation. At tree level  these
observables are equal zero. In the framework of SM the nonzero contribution is caused by final state
interaction effect. Though the effect is strongly suppressed it can be real obstacle in a new physics
searches. In this paper SM contribution to the transverse positron polarization due to the final state interaction
is considered

There are many ways to
write general effective lagrangian that describes the decay $\mu^+ \to e^+ \bar {\nu_{\mu}} \nu_e$.
In our paper the following form of the effective lagrangian will be used
\begin{eqnarray}
L_{eff}= - \frac {4 G_{f}} {\sqrt 2} \sum_{\gamma, \xi, \eta, n, m} g^{\gamma}_{\xi \eta} ~
 \bar {\mu }_{\xi}  \Gamma^{\gamma} (\nu_{\mu})_n \cdot (\bar {\nu_e})_{m}
 \Gamma_{\gamma} e_{\eta},
\label {lagr}
\end{eqnarray}
where the following designations are used: $\gamma= \mbox{scalar(S)}, \mbox{vector(V)},
\mbox{tensor(T)}$ the type of interactions, $\xi, \eta, n, m$ are the chiral projections of
spinors(left-handed(L), right-handed(R)). It should be noted that the chiral projections
of neutrinos $m, n$ are uniquely determined if $\xi, \eta$ are given. Usually
lagrangian (\ref {lagr}) parameterizes all possible SM extensions.
We imply here in the frames of SM that  only one constant $g_{LL}^V$ equals unity and all others
are zero. The deviation from this form is caused by radiative corrections.
It is shown below that one loop radiative correction leading to nonzero
transverse positron polarization can be parameterized by formula (\ref {lagr}).

The differential decay $\mu^+ \to e^+ \bar {\nu_{\mu}} \nu_e$ probability in the
framework of lagrangian (\ref{lagr}) is given by the formula\cite{Eidelman:2004wy}
\begin{eqnarray}
\frac { d^2 \Gamma} {dx dcos \theta} = \frac {m_{\mu}} {4 \pi^3} W^4_{e \mu} G_f^2
\sqrt{x^2-x_0^2} (F_{IS} (x) + P_{\mu} cos \theta F_{AS}(x)) \biggl ( 1 + {\bf s_e}
( P_{T1} {\bf e_1} + P_{T2} {\bf e_2} + P_{L} {\bf e_3} )
\biggr ),
\end{eqnarray}
where $W_{e \mu}=( m_{\mu}^2 + m_e^2 )/2 m_{\mu}$, $x=E_e/W_{e \mu}$, $x_0=m_e/W_{e \mu}$,
$P_{\mu}$ is the muon polarization, $\theta$ is the angle between muon polarization
and direction of positron momentum, $s_e$ is the unit vector in the direction
of positron spin, $P_{T1}, P_{T2}, P_{L}$ are the polarizations of the positron
corresponding to the unit vectors
\begin{eqnarray}
\bf e_3 = \frac {p_e} {|p_e|}, ~~ \bf e_2 = \frac {e_3 \times P_{\mu}} {|e_3 \times P_{\mu}|}, ~~ e_1= e_2 \times e_3
\end{eqnarray}
The functions $F_{IS}, F_{AS}, P_{T1}, P_{T2}, P_{L}$ can be expressed through the
parameters $\rho, \eta, \xi, \delta, etc.$\cite{Michel:1949qe}. In turn this parameters are functions of
coupling constants $g^{\gamma}_{\xi \eta}$.

If $g_{LL}^V=1$ and all other constants $g^{\gamma}_{\xi \eta}$ are equal zero,
the functions $F_{IS}, F_{AS}$ have the form
\begin{eqnarray}
\nonumber
F_{IS} (x) = x(1-x)+ \frac 1 6 ( 4 x^2 -3 x -x_0^2 ) \\
F_{AS} (x) = \frac 1 3 \sqrt {x^2 - x_0^2} \biggl (
1-x+ \frac 1 2 (4 x -3 + ( \sqrt { 1 - x_0^2 } - 1 ) )
\biggr )
\end{eqnarray}
In addition to this functions one needs only the expression for transverse positron polarization
$P_{T2}$ that is sensitive to $CP$-violation in muon decay. If we suppose that
all coupling constants except $g_{LL}^V$ is much less than unity and omit all terms
of second order in coupling constants than the expression for $P_{T2}$ can be written as follows
\begin{eqnarray}
P_{T2} = \frac {P_{\mu} sin \theta \cdot F_{T2} (x)}  { F_{IS} + P_{\mu} cos \theta F_{AS} (x)  },
\label{PT}
\end{eqnarray}
where $F_{T2}$ is given by
\begin{eqnarray}
F_{T2} = \frac 1 3 \sqrt {x^2 - x_0^2} \biggl (
\frac {\beta'} 8 \sqrt {1-x_0^2}
\biggr )
\end{eqnarray}
The constant $\beta'$\cite{kinoshita} can be expressed through coupling constants
\begin{eqnarray}
\beta' = 4 Im ( g_{RR}^V g_{LL}^{S*} - g_{LL}^V g_{RR}^{S*} )=  4 Im (g_{RR}^{S})
\end{eqnarray}
The last formula shows that radiative corrections lead to nonzero transverse polarization
only if the constant $g^{S}_{RR}$ acquires nonzero phase.
Thus, among all one loop correction diagrams one should consider  only those, which contain an imaginary
part. Moreover, there is no need to calculate full expression for radiative corrections since
only imaginary parts are needed. Calculating imaginary parts of one loop diagrams
we use unitarity of $S$-matrix  in the form
\begin{eqnarray}
Im T_{fi} = \frac 1 2 \sum_n T_{fn} T_{ni}^*
\label{im}
\end{eqnarray}
In many processes nonzero transverse polarization is caused by electromagnetic final state interaction\cite{Braguta:2002gz, Rogalev:2001wx, Braguta:2001nz, Okun:1967ww}.
But it is not the case for $\mu^+ \to e^+ \bar {\nu_{\mu}} \nu_e$ decay where QED corrections do not lead to nonzero
effect. It is easy to prove the statement using formula (\ref{im}). First it should be noticed that since
muon decay is considered  the intermediate particles denoted by $n$ must be lighter than muon or
one gets the amplitude($T_{ni}$) for the process in which muon decays into a number of particle with center of
mass energy greater than muon mass. Thus  positron, some electron positron pairs,
photons can be in the intermediate state and of course $\bar {\nu_{\mu}} \nu_e$. Taking into the account that in
QED $\bar {\nu_{\mu}} \nu_e$ do not interact with other intermediate particles one gets the amplitude
of the process($T_{fn}$) where positron, electron positron pairs and photons are in the intermediate state $n$ and
one positron in the final state. In other words one positron absorbs many particles what
are forbidden by the energy conservation law. This proves that in QED $Im T_{fi}$ is zero.

If one considers weak interaction in addition to electromagnetic interaction, then
 there appears diagrams that potentially
give nonzero contribution to the transverse positron polarization. These diagrams are presented in Fig. 1
Using (\ref{im}) one can write imaginary part for the diagram depicted in Fig. 1a.
\begin{eqnarray}
Im T_{fi}=- \frac {G_f} {\sqrt 2} \biggl (
\frac {G_f} {\sqrt 2} \int d \tau_2 ~ \bar {\mu} (1+ \gamma_5) \gamma_{\alpha} \hat k_1 \gamma_{\sigma} \nu_{\mu} \cdot
\bar {\nu_e} (1+ \gamma_5) \gamma^{\sigma} \hat k_2 \gamma^{\alpha} e
\biggr ),
\label{1}
\end{eqnarray}
where $k_1, k_2$-- 4-momentum of intermediate
neutrinos $\bar {\nu_{\mu}}, \nu_e$, $d \tau_2$-- two particle phase space.
The expression (\ref{1}) can be simplified by using of the formula
\begin{eqnarray}
\gamma^{\mu} \gamma^{\nu} \gamma^{\lambda} = - i \epsilon^{\mu \nu \lambda \rho} \gamma_5 \gamma_{\rho}
+ g^{\mu \nu} \gamma^{\lambda} - g^{\mu \lambda} \gamma^{\nu} + g^{\lambda \nu} \gamma^{\mu}
\label{gamma}
\end{eqnarray}
Then we get
\begin{eqnarray}
Im T_{fi}= \frac {G_f} {\sqrt 2} \biggl (
\frac {G_f} {\sqrt 2} \int d \tau_2 ~ k_1^{\rho} k_2^{\sigma}  \bar {\mu} (1+ \gamma_5) \gamma_{\rho}  \nu_{\mu} \cdot
\bar {\nu_e} (1+ \gamma_5) \gamma_{\sigma} e
\biggr ),
\label{1}
\end{eqnarray}
The integration of $k_1^{\rho} k_2^{\sigma}$ over $d \tau_2$ results in tensor structures:
$g^{\mu \nu}$ and $(p_{\nu_e}^{\mu}+p_{\nu_{\mu}}^{\mu} ) (p_{\nu_e}^{\nu}+p_{\nu_{\mu}}^{\nu} )$.
The former tensor structure gives contribution of the form $\sim  \bar {\mu} (1+ \gamma_5) \gamma^{\rho}  \nu_{\mu} \cdot
\bar {\nu_e} (1+ \gamma_5) \gamma_{\rho} e$. It is easy to see that this term gives zero effect
since it changes the phase of the coupling constant $g_{LL}^V$. The latter tensor structure
gives contribution of the form $\sim  \bar {\mu} (1+ \gamma_5) \hat {p}_{\nu_e}  \nu_{\mu} \cdot
\bar {\nu_e} (1+ \gamma_5) \hat {p}_{\nu_{\mu}} e$. Direct calculation of this contribution
to the transverse positron polarization gives zero effect. So diagram in fig 1a
gives zero transverse polarization.

Let's consider the diagram in fig. 1b. Imaginary part of the diagram can be written in the form
\begin{eqnarray}
Im T_{fi}= \frac {G_f} {\sqrt 2} \biggl (
\sqrt 2 G_f  \int d \tau_2 ~ \bar {\mu} (1+ \gamma_5) \gamma_{\alpha} \hat k_1 \gamma_{\sigma} \nu_{\mu} \cdot
\bar {\nu_e} (1+ \gamma_5)
\bigl (
(- \frac 1 2 + sin^2 \theta_W ) \gamma^{\alpha} \hat k_2 \gamma^{\sigma} +
sin^2 \theta_W \gamma^{\alpha} \gamma^{\sigma} \hat {p_e}
\bigr ) e
\biggr ),
\label{2}
\end{eqnarray}
where $\theta_W$ is Weinberg angle, $k_1, k_2$ are the 4-momentums of intermediate $\nu_{\mu}, e^+$ respectively,
$p_e$ is the final positron momentum.
Taking into the account (\ref{gamma}) one can show that formula (\ref{2})  is proportional to
$\sim  \bar {\mu} (1+ \gamma_5) \gamma^{\rho}  \nu_{\mu} \cdot
\bar {\nu_e} (1+ \gamma_5) \gamma_{\rho} e$ what again gives zero contribution to transverse polarization.

First nonzero contribution comes from the diagram in fig. 1c. After necessary transformations  the imaginary part of
this diagram can be represented as follows
\begin{eqnarray}
Im T_{fi}=- \frac {G_f} {\sqrt 2} \biggl (
\sqrt 2 G_f  \int d \tau_2 ~ \bar {\mu} (1+ \gamma_5) \gamma_{\alpha}  \nu_{\mu} \cdot
\bar {\nu_e} (1+ \gamma_5)
\bigl (
 (1 - 2 sin^2 \theta_W ) \hat k_1  \gamma^{\alpha} \hat k_2 - 4 m_e sin^2 \theta_W k_2^{\alpha}
\bigr ) e
\biggr ),
\label{3}
\end{eqnarray}
here $k_1, k_2$ are the 4-momentums of intermediate $e^+, \nu_{e}$ respectively.
The integration over $\tau_2$ can be carried out using the formulae
\begin{eqnarray}
\nonumber
\int d \tau_2 k_2^{\alpha} &=& 4 \pi \tau_2^2 P^{\alpha} \\
\int d \tau_2 k_1^{\alpha} k_2^{\beta} &=& \frac {2 \pi} 3 (P^2-m_e^2) \tau_2^2 g^{\alpha \beta} +
\frac {4 \pi} 3 (P^2 + 2 m_e^2) \tau_2^2 \frac {P^{\alpha} P^{\beta} } {P^2},
\end{eqnarray}
where $P=p_e +p_{\nu_e}$, $\tau_2 = (P^2-m_e^2)/8 \pi P^2 $ is $(e^+, \nu_e)$ phase space.
Having made necessary transformations one gets the result
\begin{eqnarray}
Im T_{fi} =- \frac {G_f} {\sqrt 2} 8 \sqrt 2 \pi G_f  m_e m_{\mu}  \tau_2^2 \biggl (
\frac {1} 3 (1+2 \frac {m_e^2} {P^2} ) (1-2 sin^2 \theta_W)   +  2  sin^2 \theta_W
\biggr ) \bar {\mu} (1- \gamma_5) \nu_{\mu} \cdot  \bar {\nu}_e (1+ \gamma_5) e
\end{eqnarray}
So, the diagram in fig. 1c gives nonzero contribution
to the $Im g_{RR}^S$
\begin{eqnarray}
Im g_{RR}^S =  \frac {\sqrt 2} {8 \pi}  G_f  m_e m_{\mu}  (1- \frac {m_e^2} {P^2} )^2 \biggl (
\frac {1} 3 (1+2 \frac {m_e^2} {P^2} ) (1-2 sin^2 \theta_W)   +  2  sin^2 \theta_W
\label{im1}
\biggr )
\end{eqnarray}
Omitting the terms $m_e^2 / P^2$ which are much less than unity  formula (\ref{im1}) becomes much simpler
\begin{eqnarray}
Im g_{RR}^S =  \frac {1} {12 \sqrt 2 \pi}  G_f  m_e m_{\mu}
(1+4 sin^2 \theta_W)
\end{eqnarray}
Last diagram that can give nonvanishing contribution is presented in fig 1d. Imaginary part
of the diagram has the form
\begin{eqnarray}
Im T_{fi}= - \frac {G_f} {\sqrt 2} \biggl (
\frac {G_f} {\sqrt 2}   \int d \tau_2 ~ Tr({\hat k_2} \gamma_{\alpha} {\hat k_1} \gamma_{\beta}  ) ~ \bar {\mu} (1+ \gamma_5) \gamma^{\alpha}  \nu_{\mu} \cdot
\bar {\nu_e} (1+ \gamma_5) \gamma^{\beta}  e
\biggr )
\label{4}
\end{eqnarray}
Transforming expression (\ref {4}) by the procedure which we used above, one gets:
\begin{eqnarray}
Im T_{fi} =- \frac {G_f} {\sqrt 2}
\biggl ( Im g_{RR}^S
\bar {\mu} (1- \gamma_5) \nu_{\mu} \cdot  \bar {\nu}_e (1+ \gamma_5) e
\biggr ),
\end{eqnarray}
where
\begin{eqnarray}
Im g_{RR}^S =  \frac {1} {6 \sqrt 2 \pi} G_f m_{\mu} m_e (1+2 \frac {m_e^2} {P^2}) (1- m_e^2/P^2)^2 =  \frac {1} {6 \sqrt 2 \pi} G_f m_{\mu} m_e
\end{eqnarray}
In last equality the terms $m_e^2/ P^2$ are omitted. Summing the contributions from fig 1c. and fig 1d. we get
\begin{eqnarray}
Im g_{RR}^S =   \frac {1} {12 \sqrt 2 \pi} G_f m_{\mu} m_e (3 + 4 sin^2 \theta_W)
\end{eqnarray}
Numerical estimation of this quantity gives us the result
\begin{eqnarray}
Im g_{RR}^S = 4 \times 10^{-11}
{\label{res}}
\end{eqnarray}
Now it is seen that the value obtained in the experiment\cite{Danneberg:2005xv}
\begin{eqnarray}
Im g_{RR}^S = ( 5.2 \pm 14.0 \pm 2.4 ) \times 10^{-3}
\end{eqnarray}
is much greater than the value of the effect predicted in the framework of SM.
This fact allows us to state that the search of the effect of $CP$-violation
in purely lepton decay by the measurement of transverse positron polarization
is not obscured by SM contributions. So the measurement of the
transverse positron polarization is very promising in the search of new physics
since it either discovers $CP$-violation in muon decay or puts very strict bounds on
parameters of different SM extensions.

The author thanks professor A.K. Likhoded, V.V. Bytev and  A.S. Zhemchugov for useful discussions.
This work was partially
supported by Russian Foundation of Basic Research under grant 04-02-17530, Russian Education
Ministry grant E02-31-96, CRDF grant MO-011-0, Scientific School grant SS-1303.2003.2 and
Dynasty foundation.

\newpage

\begin{figure}[ph]
\begin{picture}(150, 200)
\put(0,70){\epsfxsize=9cm \epsfbox{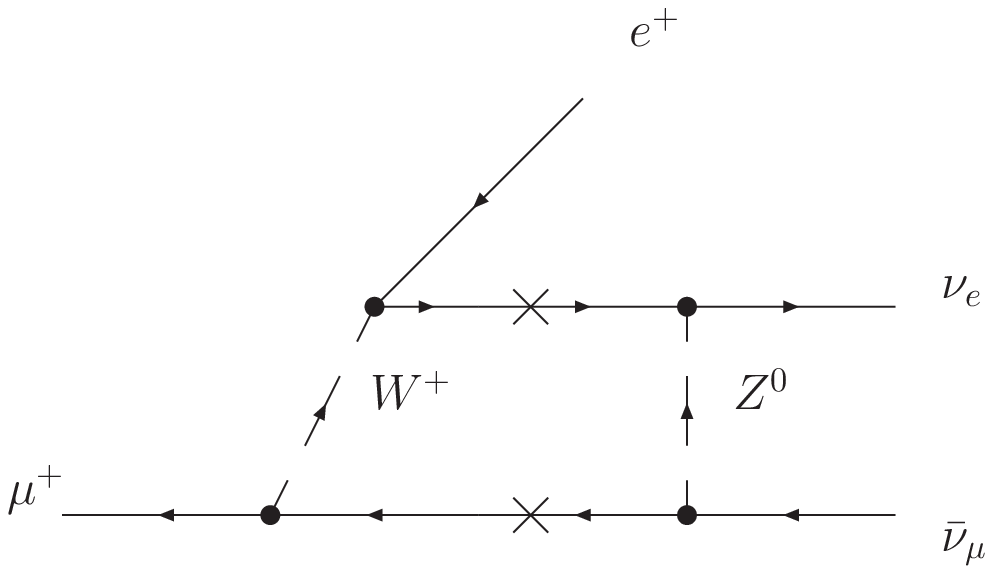}}
\put(100,60){\bf{Fig. 1a}}
\put(250,80){\epsfxsize=9cm \epsfbox{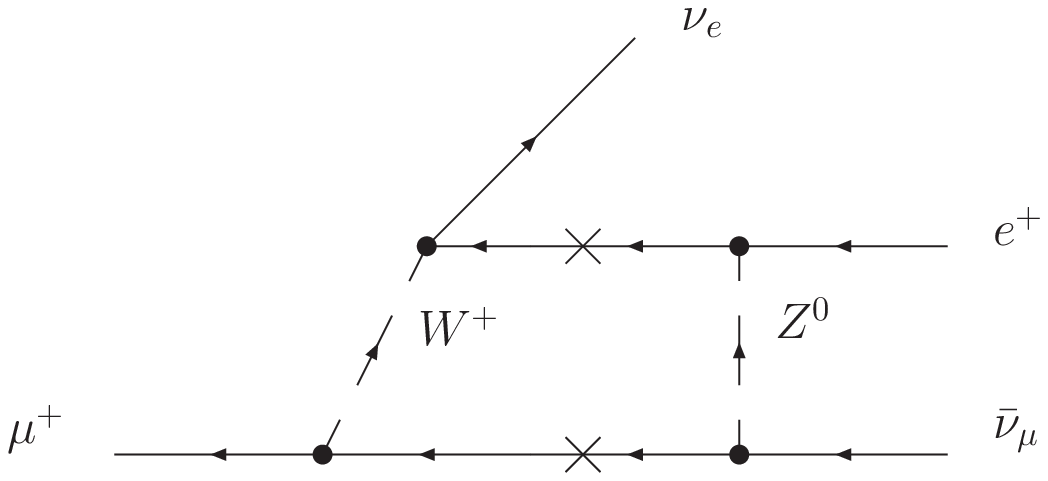}}
\put(350,60){\bf{Fig. 1b}}
\put(0,-150){\epsfxsize=9cm \epsfbox{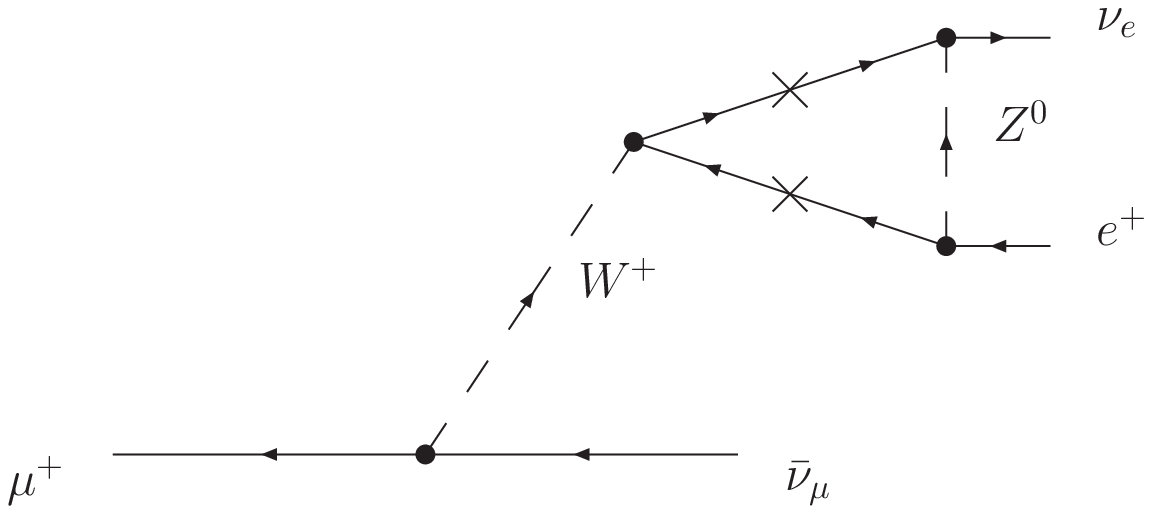}}
\put(100,-170){\bf{Fig. 1c}}
\put(250,-150){\epsfxsize=9cm \epsfbox{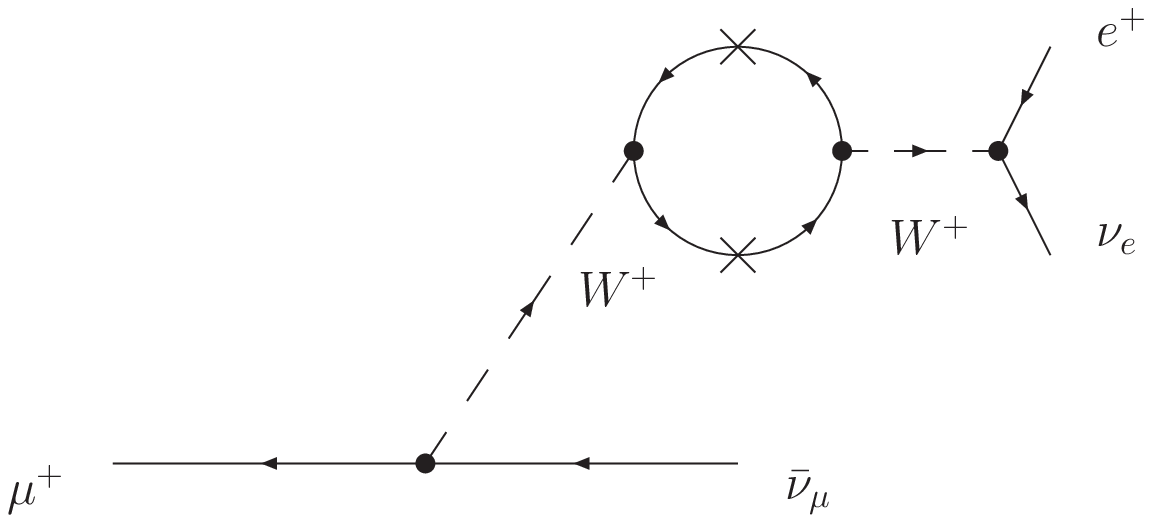}}
\put(350,-170){\bf{Fig. 1d}}
\end{picture}
\end{figure}

\end{document}